\documentclass[a4paper]{article}

\usepackage{INTERSPEECH2021}
\usepackage{mathrsfs}
\usepackage{subfigure}
\usepackage{graphicx}
\usepackage{float} 
\usepackage{multirow}
\usepackage{makecell}
\title{Bidirectional Multiscale Feature Aggregation for Speaker Verification}
\name{Jiajun Qi, Wu Guo, Bin Gu}
\address{
  National Engineering Laboratory for Speech and Language Information Processing,\\
  University of Science and Technology of China, Hefei, China}
\email{jiajun97@mail.ustc.edu.cn, guowu@ustc.edu.cn, bin2801@mail.ustc.edu.cn}

\begin{document}
\maketitle
\begin{abstract}
 In this paper, we propose a novel bidirectional multiscale feature aggregation (BMFA) network with attentional fusion modules for text-independent speaker verification. The feature maps from different stages of the backbone network are iteratively combined and refined in both a bottom-up and top-down manner. Furthermore, instead of simple concatenation or element-wise addition of feature maps from different stages, an attentional fusion module is designed to compute the fusion weights. Experiments are conducted on the NIST SRE16 and VoxCeleb1 datasets. The experimental results demonstrate the effectiveness of the bidirectional aggregation strategy and show that the proposed attentional fusion module can further improve the performance.
\end{abstract}

\section{Introduction}
\label{section:intro}
Speaker verification (SV) is used to verify a person’s claimed identity from their voice characteristics. During the last decade, the i-vector \cite{dehak2010front} algorithm combined with a probabilistic linear discriminant analysis (PLDA) \cite{kenny2010bayesian} backend for similarity scoring has become very popular due to its good performance and ability to compensate for within-speaker variations.

In recent years, this paradigm has shifted to employ deep neural networks (DNNs) to extract speaker embeddings, called the d-vector \cite{variani2014deep} or x-vector \cite{snyder2018x}. DNN-based speaker embedding extraction outperforms conventional i-vectors and achieves state-of-the-art performance in many cases, especially in conditions of short-duration SV. These embedding networks typically consist of three main components: (1) several frame-level layers to obtain high-level feature representations, (2) a pooling layer that aggregates the frame-level representations across the temporal dimension and (3) a bottleneck layer that projects the pooled vector into a low-dimensional speaker embedding.

The frame-level embeddings are usually modeled by time-delay neural networks (TDNNs) \cite{zhang2017end,jiang2019effective,desplanques2020ecapa}, long short-term memory (LSTM) networks  \cite{wan2018generalized}, or convolutional neural networks (CNNs) \cite{gao2018improved,lin2020wav2spk}. Similar to the image classification field, the deep residual network (ResNet) \cite{he2016deep} has also been widely adopted as the backbone architecture of embedding networks in SV systems \cite{chung2018voxceleb2,cai2018exploring,gao2019improving} due to its good performance. ResNet usually comprises an input convolutional layer and 4 stages, where each stage represents a group of residual convolutional layers that output feature maps of the same scale.

Generally, the DNNs of the embedding extractor contain many layers. To exploit hierarchical time-frequency context information from different layers, multiscale feature aggregation (MFA) has been developed \cite{desplanques2020ecapa,gao2018improved,gao2019improving,seo2019shortcut,hajavi2019deep}. Gao et al. \cite{gao2019improving} proposed aggregating multistage feature maps with different resolutions using concatenation. Seo et al. \cite{seo2019shortcut} proposed concatenating pooled embeddings from each stage to build shortcut connections from low- and middle-layer features to the final speaker embeddings. Similar to \cite{seo2019shortcut}, Hajavi et al. \cite{hajavi2019deep} used weighted summation instead of concatenation on pooled embeddings from multiple stages for aggregation. These MFA architectures extract features from different layers and aggregate them with elementwise addition or concatenation operations. More recently, a progressive MFA method \cite{jung2020improving} was introduced to create multiscale features that have high-level speaker information at all layers by applying feature pyramid module (FPM), which consists of a top-down pathway and lateral connection.

Existing MFA methods have achieved outstanding performance on SV tasks; however, there is still room for further improvement of MFA. First, the abovementioned works aggregate the multistage features only through a unidirectional pathway, and this setup may limit the feature interaction of lower and higher layers. Second, concatenation and elementwise addition are usually employed in feature aggregation steps, and these simple operations have limited abilities to extract relevant information from different features due to their fixed fusion weights.

In this work, we propose a novel bidirectional multiscale feature aggregation (BMFA) network with attentional fusion modules (AFM) for speaker verification. For the proposed method, features of different stages are first extracted from the ResNet backbone. Then, a bidirectional feature aggregation strategy, which consists of top-down and bottom-up pathways, is used to progressively fuse these features. Additionally, inspired by \cite{dai2021attentional}, an attentional fusion module is employed to learn better fusion weights. We evaluate our experiments on the NIST SRE16 \cite{sadjadi20172016} and VoxCeleb1 \cite{nagrani2017voxceleb} datasets. The experimental results show that the proposed BMFA can achieve better performance than the existing MFA strategy and obtain further improvement when combined with the proposed attentional fusion module.

The rest of the paper is organized as follows. Section 2 reviews the existing architecture of MFA networks. Section 3 describes the proposed bidirectional model and attentional fusion module in detail. Then, the experimental setup is presented in section 4. The results and the analysis are presented in section 5. Finally, the conclusions are given in section 6.

\section{Multiscale Feature Aggregation}
\label{section:MFA}
In this section, we briefly review different MFA methods in SV field. The architecture of conventional MFA methods \cite{gao2019improving,seo2019shortcut,hajavi2019deep} is illustrated in Fig. 1 (a). Specifically, the feature maps of multiple scales are extracted from different stages in the ResNet-based network and then aggregated simultaneously either before or after temporal pooling. Concatenation or weighted addition is usually employed as the fusion method.

A progressive top-down MFA approach using FPM is proposed in \cite{jung2020improving}, as shown in Fig. 1 (b). The FPMs iteratively fuses features extracted from different stages in the backbone network, where the top-down pathway upsamples the feature map from the higher pyramid level to produce high-resolution features and then enhances them with features from the backbone network by elementwise addition. Each fused feature is filtered by a convolutional layer and pooled by a temporal pooling layer to generate embeddings at different pyramid stages. Then, these embeddings are concatenated to obtain more discriminative speaker embedding.

\begin{figure}[t]
  \centering
  \subfigure[]{
  \includegraphics[height=6.1cm]{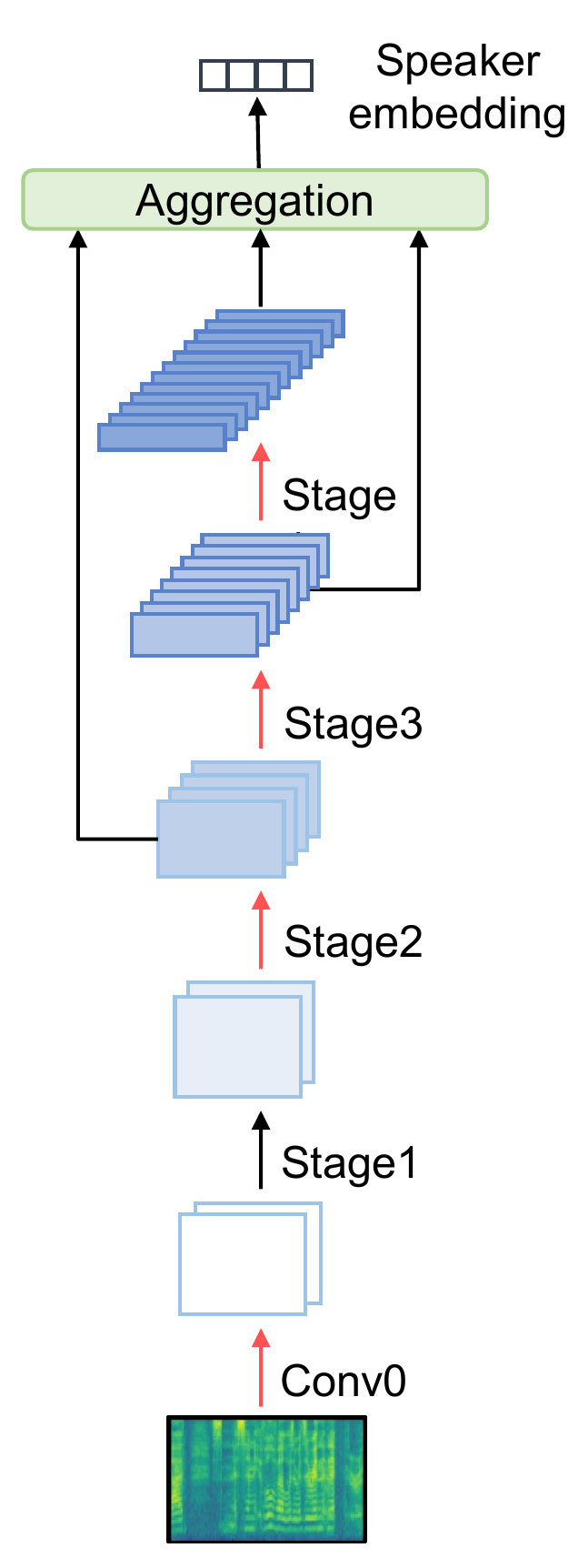}}
  \hspace{-0.cm}
  \subfigure[]{
  \includegraphics[height=6.1cm]{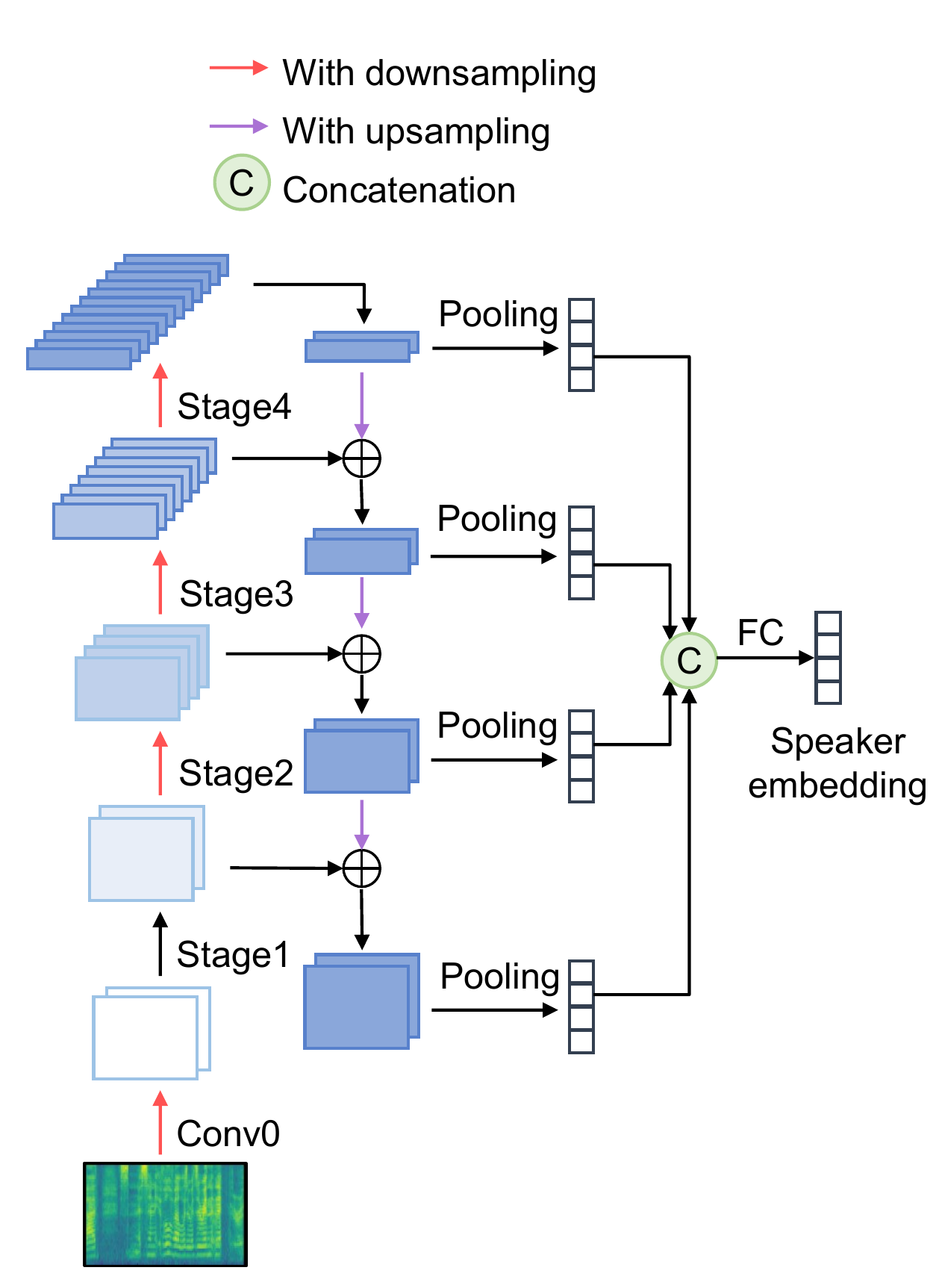}}
  \caption{ Illustration of different multiscale aggregation architectures in SV. (a) Feature aggregation through concatenation or addition, (b) top-down feature aggregation with FPM}
  \label{fig:MFA}
\end{figure}

\begin{table}[bh]
\footnotesize
  \caption{The architecture of the backbone ResNet34. Inside the brackets is the shape of a residual block, and outside the brackets is the number of stacked blocks on a stage. The input size is T×64.}
  \label{tab:ResNet34}
  \centering
  \begin{tabular}{|c|c|c|c|}
    \hline
    Stage & Layer name  &   Output size             &   ResNet34                   \\
    \hline
    - & Conv0       &   $T\times64\times32$     &   $7\times7$,32,stride 2      \\
    \hline
    1 & Conv1\_x    &   $T/2\times32\times32$	&   $\begin{bmatrix} 3\times3,32\\3\times3,32\\\end{bmatrix}\times3$   \\
    \hline
    2 & Conv2\_x	&   $T/2\times16\times64$	&   $\begin{bmatrix} 3\times3,32\\3\times3,32\\\end{bmatrix}\times4$  \\
    \hline
    3 & Conv3\_x	&   $T/2\times8\times128$   &	$\begin{bmatrix} 3\times3,32\\3\times3,32\\\end{bmatrix}\times6$   \\
    \hline
    4 & Conv4\_x	&   $T/2\times4\times256$   &	$\begin{bmatrix} 3\times3,32\\ 3\times3,32\\\end{bmatrix}\times3$   \\
    \hline
  \end{tabular}
\end{table}

\begin{figure}[t]
  \centering
  \subfigure[]{
  \includegraphics[height=6cm]{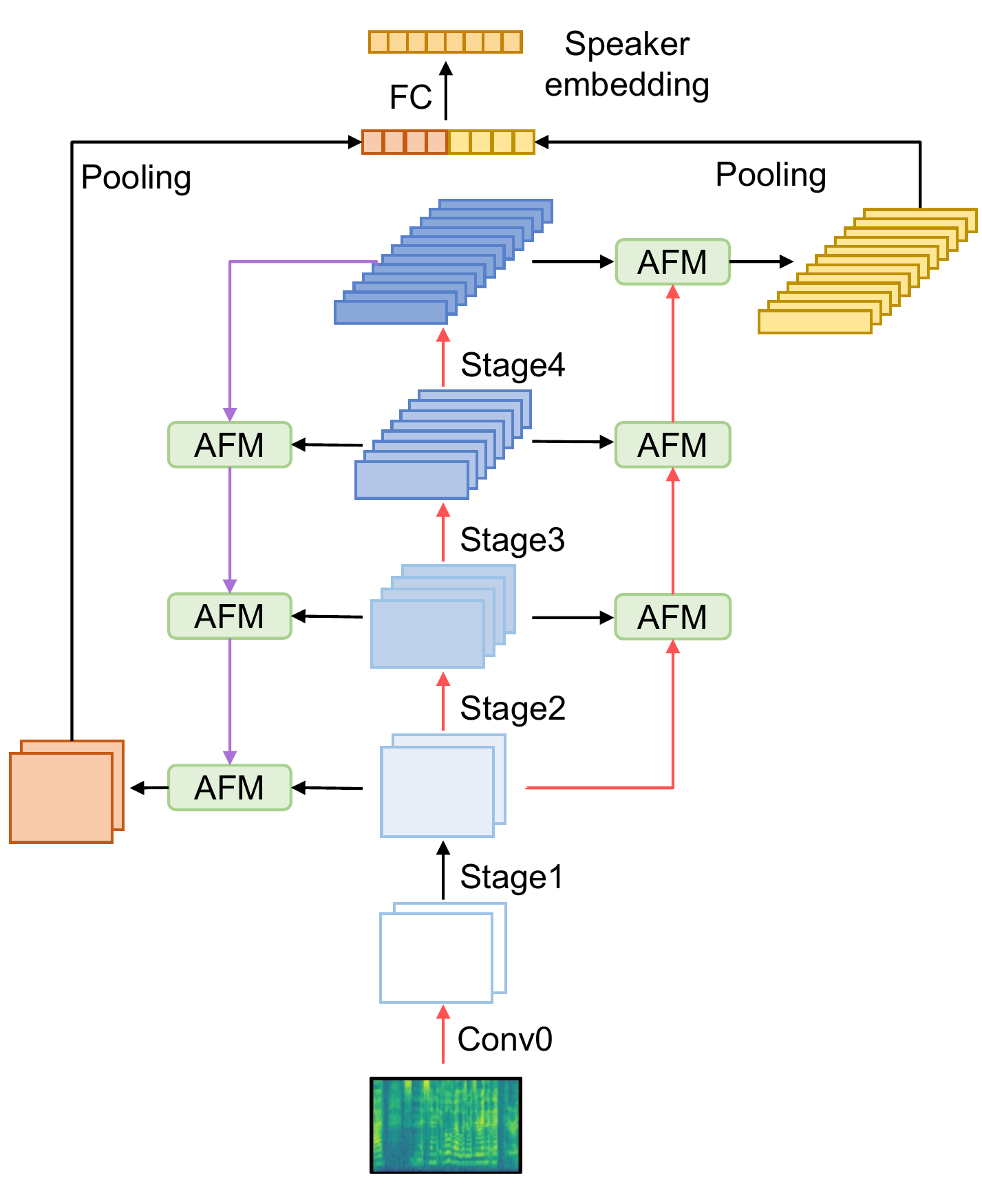}}
  \hspace{-0.3cm}
  \subfigure[]{
  \includegraphics[height=5.43cm]{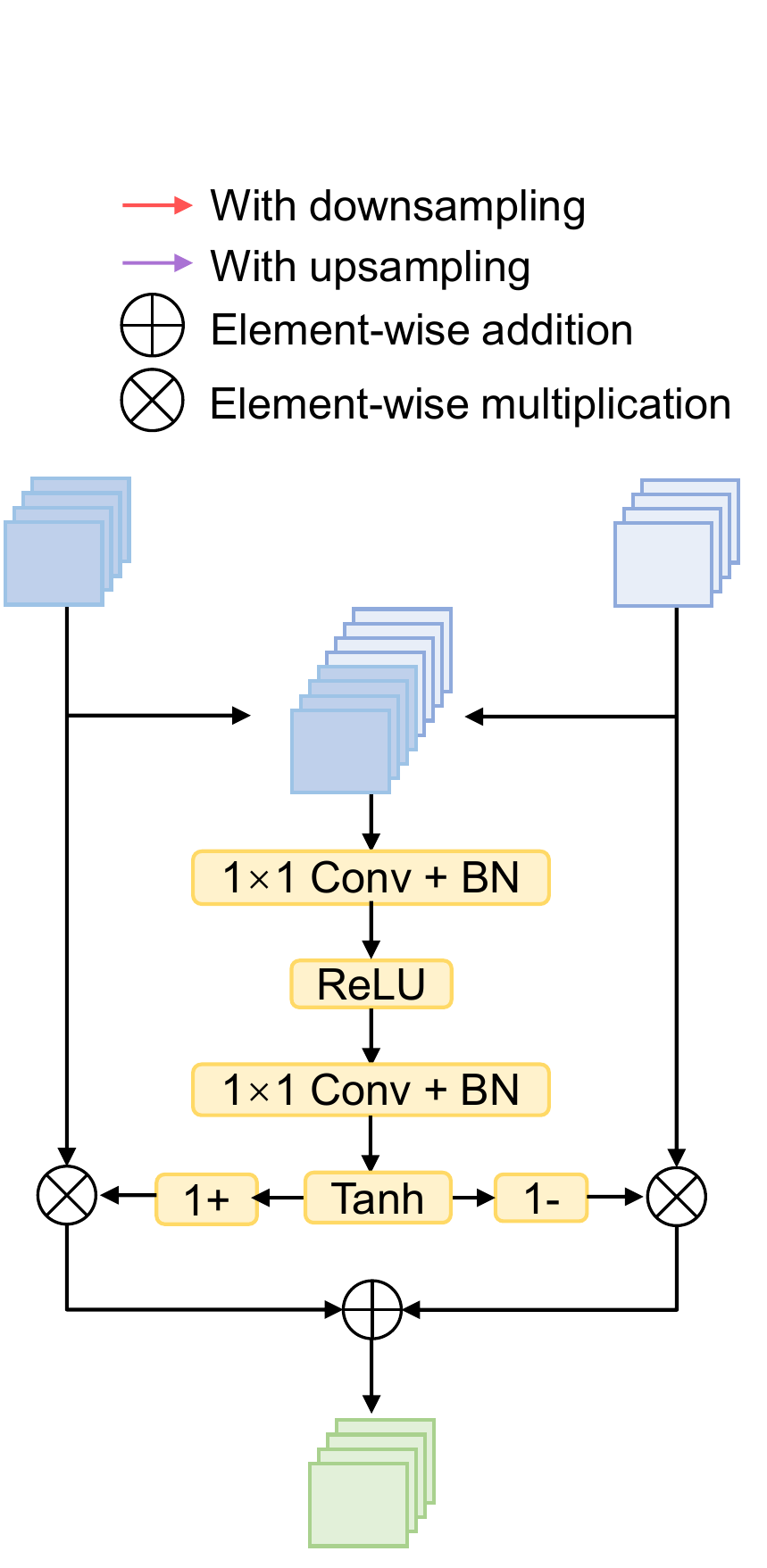}}
  \caption{ (a) Illustration of the proposed bidirectional multiscale aggregation architectures. (b)The structure of the attentional fusion module}
  \label{fig:BMFA}
\end{figure}

\section{Proposed Method}
\label{section:proposed_method}
In this section, we discuss details of the proposed bidirectional multiscale feature aggregation (BMFA) network along with the attentional fusion module. 

\subsection{Bidirectional Multiscale Feature Aggregation}
\label{subsection:BMAF}
Fig. 2 (a) schematically depicts the overall architecture of the proposed network. The network consists of three branches: a backbone branch, a bottom-up aggregation branch, and a top-down aggregation branch.

 \textbf{Backbone branch:} We utilize ResNet34 as the backbone model in this work. The detailed configurations of ResNet34 are listed in Table 1. We extract feature maps with different resolutions from the last layer of every stage and denote the feature map of the $i$-th stage as $\bm{C}_{i}$ for $i$=1,2,3,4. Then these features are transmitted to both top-down and bottom-up aggregation branches separately. 
 
 \textbf{Top-down aggregation branch:} By propagating high-level semantic context from top layers into bottom layers, top-down aggregation can help to suppress noisy information in bottom layers. Specifically, each feature map ($\bm{F}_{i}^{tb},i=3,2,1$) of the top-down branch is iteratively built by combining the backbone feature map ($\bm{C}_{i}$) from the same stage and the higher-stage top-down feature map ($\bm{F}_{i+1}^{tb}$):
\begin{equation}
    \begin{cases}
        \bm{F}^{tb}_4=\bm{C}_4 \\
        \bm{F}^{tb}_i=AFM(\bm{\mathcal{U}}(\bm{\mathcal{B}}(\bm{W}^{tb}_i)*\bm{F}^{tb}_{i+i}),\bm{\mathcal{B}}(\bm{W}^{lc_{tb}}_i*\bm{C}_i))\\
    \end{cases}
\end{equation}
where $\bm{\mathcal{B}}(\cdot)$ denotes batch normalization \cite{ioffe2015batch} and $\bm{\mathcal{U}}(\cdot)$ is the upsampling operation in the frequency dimension with a factor of 2. $*$ denotes the convolution operation, and $AFM(\cdot)$ is operator of the proposed attentional fusion module (see section 3.2). $\bm{W}_i^{tb}$ and $\bm{W}_i^{lc}$ are both $1\times1$ convolution filters, where $\bm{W}_i^{tb}$ is used to reduce the channel number to be same with $\bm{C}_i$, while $\bm{W}_i^{lc_{tb}}$ is used in lateral connection to re-encode the channel information of feature map $\bm{C}_i$. Additionally, we append a $3\times3$ convolutional layer on the bottom feature map $\bm{F}_1^{tb}$ to refine the features, followed by statistical pooling layer to encapsulate the variable size features into a fixed size embedding $\bm{h}^{tb}$. 

\textbf{Bottom-up aggregation branch:} In the bottom-up branch, the complementary low-level details from the bottom layers are fused into high-level features. Specifically, each feature map ($\bm{F}_i^{bt},i=2,3,4$) of the bottom-up branch is iteratively built by combining the backbone feature map ($\bm{C}_i$) from the same stage and the lower-stage bottom-up feature map ($\bm{F}_{i-1}^{bt}$):
\begin{equation}
   \begin{cases}
\bm{F}^{bt}_1=\bm{C}_1 \\
\bm{F}^{bt}_i=AFM(\bm{\mathcal{B}}(\bm{W}^{bt}_i*\bm{\mathcal{D}}(\bm{F}^{bt}_{i-i})),\bm{\mathcal{B}}(\bm{W}^{lc_{bt}}_i*\bm{C}_i))\\
\end{cases} 
\end{equation}
where $\bm{\mathcal{D}}(\cdot)$ denotes the downsampling operation in the frequency dimension with a factor of 2. $\bm{W}_i^{bt}$ and $\bm{W}_i^{lc_{bt}}$ are both $1\times1$ convolution filters, where $\bm{W}_i^{tb}$ is used to increase the channel number to be same with $\bm{C}_i$, and $\bm{W}_i^{lc_{bt}}$ has the same purpose as $\bm{W}_i^{lc_{tb}}$. Similar to the top-down branch, we append a $3\times3$ convolutional layer on the top feature map $\bm{F}_1^{bt}$, followed by a statistic pooling layer to generate utterance-level bottom-up embedding $\bm{h}^{bt}$. 

Finally, we perform a fully connected layer on the concatenation of $\bm{h}^{bt}$ and $\bm{h}^{tb}$ to achieve a nonlinear transformation and dimension reduction to obtain speaker embedding for verification. In the training phase, an additional fully connected layer and softmax layer are appended to the speaker embeddings for the speaker classification task.

\subsection{Attentional Fusion Module}
\label{subsection:AFM}
Existing MFA architectures rarely focus on feature fusion modules and usually employ simple operations such as elementwise addition or concatenation, which are suboptimal for combining feature maps of different importance. To address this problem, we design an attentional fusion module (AFM) to compute attention weights.

As shown in Figure 2 (b), AFM takes the concatenation of two feature maps as the input and computes the time-frequency attention map using two convolutional layers. Given two feature maps $\bm{X}$,$\bm{Y}$ with $C$ channels, the attention map $\bm{S}$ is computed by:
\begin{equation}
    \bm{S}=tanh(\bm{\mathcal{B}}((\bm{W}_2*ReLU(\bm{B}(\bm{W}_1*[\bm{X},\bm{Y}])))))
\end{equation}
where $[\cdot]$ refers to the concatenation along the channel dimension. $\bm{W}_1$ and $\bm{W}_2$ are $1\times1$ convolution filters for aggregating local channel context information with output channel sizes of $C/r$ and $C$, respectively. $r$ is the channel reduction ratio (we set $r$=4 in this work). Note that $\bm{S}$ has the same shape as $\bm{X}$ and $\bm{Y}$.
Then, we fuse feature maps $\bm{X}$ and $\bm{Y}$ using the attention map as:
\begin{equation}
    \bm{F}=AFM(\bm{X},\bm{Y})=(1\oplus{}\bm{S})\otimes{}\bm{X}+(1\ominus{}\bm{S})\otimes{}\bm{Y}
\end{equation}
where $\oplus{}$ and $\ominus{}$ denote broadcasting addition and subtraction, respectively. $\otimes{}$ denotes elementwise multiplication. In AFM, $\bm{S}$ represents the correlation between inputs $\bm{X}$ and $\bm{Y}$. After broadcasting addition and subtraction on $\bm{S}$, the fusion weights of $\bm{X}$ and $\bm{Y}$ are complementary to each other, which enables module to conduct a soft aggregation of $\bm{X}$ and $\bm{Y}$ by enhancing one and suppressing the other.

\section{Experimental Setup}
\label{sec:exp}
\subsection{Datasets and evaluation metrics}
The experiments are carried out on the NIST SRE16 evaluation dataset, and the VoxCeleb1 dataset. For the NIST SRE16 dataset, the training data mainly consist of telephone speech (with a small amount of microphone speech) from the NIST SRE2004-2010, Mixer 6 and Switchboard datasets. We also use the data augmentation techniques described in \cite{snyder2018x}, which employ the babble, music, noise and reverb augmented data to increase the quantity and diversity of the existing training data. In summary, there were a total of 183,457 recordings from 7,001 speakers, including approximately 96,000 randomly selected augmented recordings. The performance is evaluated in terms of equal error rate (EER), the minimal detection cost function (minDCF) and actual detection cost function (actDCF) calculated using the SRE16 official scoring software. For the NIST SRE16, the equalized results are used.

For the VoxCeleb1 test set, we use the development part of the VoxCeleb2 dataset \cite{chung2018voxceleb2} without any data augmentation as training data, which contains 5994 speakers with 1,092,009 utterances. The performance is evaluated in terms of EER and minDCF with $P_{target}=10^{-2}$.

\subsection{Input features}
The input acoustic features are 64-dimensional FBanks with a frame length of 25 ms that are mean-normalized over a sliding window of up to 3 s. An energy-based voice activity detector (VAD) is used to remove all the silent frames. The utterances are cut into chunks with durations ranging from 200 to 400 frames for training. 

\subsection{Model configuration}
In this work, we implement all models in TensorFlow toolkit \cite{abadi2016tensorflow}. The other procedures (including data processing, feature extraction and the PLDA backend) are implemented using the Kaldi toolkit \cite{povey2011kaldi}. The network is optimized using the Adam optimizer with minibatch size of 64, and the learning rate gradually decreases from 1e-3 to 1e-4. We use bilinear interpolation to upsample the feature maps in the top-down branch and 3$\times$3 convolution with a stride of 2 in the frequency dimension to downsample the feature maps in the bottom-up branch. Both fully connected layers after the statistic pooling layer have 512 nodes. The additive margin softmax (AM-softmax) \cite{liu2017sphereface} is used as the training criterion for all models, and the margin $m$ and scaling factor $s$ are set to 0.15 and 30, respectively. We build four systems for comparison, and the configurations of each system are listed as follows:
\begin{enumerate}
\item Baseline: ResNet34 network.
\item MFA\_S34: ResNet34 network with MFA as described in \cite{gao2019improving} (Fig.1 (a)), aggregating features of stage 3 and stage 4.
\item MEA\_FPM: ResNet34 network with multiscale embedding aggregation (MEA) using FPM as described in \cite{jung2020improving} (Fig.1 (b)).
\item BMFA-AFM: the proposed network in this paper, i.e., the ResNet34 network with bidirectional multiscale feature aggregation (BMFA) using the attentional fusion module (AFM) as described in section 3 (Fig. 2 (a)).
\end{enumerate}

\subsection{PLDA Backend}
For the NIST SRE16, the DNN embeddings are centered using the unlabeled development data and are projected using LDA, which reduces the dimensions of the speaker embedding to 100. The PLDA model is trained and then adapted to the unlabeled data through unsupervised adaptation in Kaldi. For VoxCeleb1, For VoxCeleb1, no adaptation technique or LDA projection is used.

\begin{table*}[htbp]
  \caption{Comparison results of different multiscale feature aggregation strategies and different fusion methods on the NIST SRE16. For the fusion module, Add and Concat denote elementwise addition and concatenation operation, respectively. }
  \label{tab:SRE16}
  \centering
  \begin{tabular}{ccccccccccc}
    \hline
 Aggregation  &  Fusion    & &  Pooled  & & & Cantonese & & &Tagalog\\
   \cline{4-4} \cline{7-7} \cline{10-10}
    strategy    &  method    & EER(\%)  & minDCF & actDCF & EER(\%)  & minDCF & actDCF & EER(\%)   & minDCF & actDCF \\
    \hline
    Baseline    &    -       & 6.54 & 0.5505 & 0.5671 & 2.99 & 0.3017 & 0.3225 & 10.05 & 0.7357 & 0.8118    \\
    \hline
    MFA\_S34	& Concat \cite{gao2019improving} & 6.34 & 0.5348 & 0.5482 & 2.87 & \textbf{0.2897} & \textbf{0.3036} & 9.84  & 0.7218 & 0.7928    \\
    MFA\_S34    & AFM        & \textbf{6.21} & \textbf{0.5118} & \textbf{0.5191} & \textbf{2.70} & 0.3033 & 0.3215 & \textbf{9.66}  & \textbf{0.6889} & \textbf{0.7166}    \\
    \hline
    MEA\_FPM	& Add \cite{jung2020improving}   & 6.33 & 0.5031 & 0.5082 & 2.53 & 0.2832 & \textbf{0.3000} & 10.06 & 0.6963 & 0.7164    \\
    MEA\_FPM	& AFM        & \textbf{6.10} & \textbf{0.4926} & \textbf{0.4981} & \textbf{2.50} & \textbf{0.2786} & 0.3022  &  \textbf{9.61} &  \textbf{0.6757} &  \textbf{0.6939}    \\
    \hline
    BMFA	    &Concat	     & 6.23 & 0.5001 & 0.5037 & 2.66 & 0.2694 & 0.2917 & 9.82 & 0.6952 & 0.7158    \\
    BMFA        &Add         & 6.00 & 0.4964 & 0.5017 & 2.51 & 0.2797 & 0.2938 & 9.63 & 0.6856 & 0.7096    \\
    BMFA        &AFM         & \textbf{5.79} & \textbf{0.4620} & \textbf{0.4673} & \textbf{2.39} & \textbf{0.2531} & \textbf{0.2843} & \textbf{9.20} & \textbf{0.6438} & \textbf{0.6503}    \\
    \hline
  \end{tabular}
    \vspace{-0.5cm}
\end{table*}

\section{Results and Analysis}
\subsection{Results on the SRE16 dataset}
\subsubsection{Comparison between different aggregation strategies }
Table 2 presents the performance of the systems with different feature aggregation strategies on the NIST SRE16 dataset. Furthermore, we compare the proposed attentional fusion module (AFM) with the conventional elementwise addition and concatenation fusion methods in Table 2.

It can be observed that the systems with the multiscale aggregation strategy outperform the ResNet34 baseline system in all evaluation conditions.Additionally, we observe that the proposed attentional fusion module (AFM) can obtain consistent performance improvement over the conventional fusion methods for almost all three aggregation strategies. Among all of the abovementioned aggregation strategies, our BMFA-AFM system, employing bidirectional feature aggregation and attentional fusion modules, achieves the best performance, which improves the EER, minDCF and actDCF over the baseline system on the pooled set by 11.5\%, 16.1\% and 17.6\%, respectively.

\subsubsection{Ablation study }
To investigate the effect of (1) the bidirectional aggregation strategy and (2) the features from different stages of the backbone network for the proposed BMFA-AFM model, we perform an ablation study, as shown in Table 3. For simplicity, only the SRE16 pooled results are listed. 
In Table 3, MFA\_Top-down, MFA\_Bottom-up denote multiscale feature aggregation using features from stage 1 to stage 4 with only top-down or bottom-up branch (described as section 3.1), respectively. BMFA\_S34 refers to BMFA using features from stage 3 and stage 4, and BMFA\_S234 refers to BMFA using features from stage 2, 3 and 4. BMFA\_S1234 refers to BMFA using features of stage 1 to 4, namely BMFA-AFM. All models apply AFM for feature fusion.

We can observe from the first set of experiments (‘+MFA\_Top-down’ and ‘+MFA\_Bottom-up’) that both top-down and bottom-up aggregation strategies can boost the performance over the baseline. Compared to the top-down aggregation strategy, the bottom-up strategy obtains a more obvious improvement. We can observe that combining the bottom-up and top-down aggregation (BMFA) further improves the performance by a large margin. Furthermore, the performance can be improved gradually by adding more features from lower stages for BMFA systems, as shown in the bottom three rows in Table 3. This consistent improvement indicates that features of lower stages can provide extra complementary information for the final verification task.

\subsection{Comparison with recent methods on Voxceleb1}
In this section, we compare the proposed method with several recent approaches on the VoxCeleb1 dataset. As shown in Table 5, with the same ResNet-based feature extractor, the proposed method achieves the best results among all listed methods. Compared to the recent multiscale aggregation methods (ResNenXt41-MSA and ResNet34-FPM), our model outperforms them by a large margin. Specifically, the proposed method achieves the best results, with EER of 2.98\% and minDCF of 0.3008 when training on VoxCeleb1 dev and EER of 1.73\% and minDCF of 0.1873 when training on VoxCeleb2 dev. The results demonstrate that the proposed bidirectional aggregation strategy with attentional fusion module is a more efficient method for the aggregation of multiscale features. 

\begin{table}[htbp]
  \vspace{-0.2cm}
  \caption{Ablation study for different directional aggregation and different features.}
  \label{tab:SRE16}
  \centering
  \begin{tabular}{cccc}
    \hline
   Aggregation strategy  & EER(\%)  & minDCF & actDCF \\
    \hline
    Baseline        & 6.54 & 0.5505 & 0.5671   \\
    \hline
    +MFA\_Top-down	& 6.33 & 0.5215 & 0.5296   \\
    +MFA\_Bottom-up	& 6.12 & 0.5092 & 0.5150    \\
    \hline
    +BMFA\_S34	    & 6.34 & 0.4980 & 0.5036    \\
    +BMFA\_S234	    & 6.09 & 0.4822 & 0.4911    \\
    +BMFA\_S1234	& \textbf{5.79} & \textbf{0.4620} & \textbf{0.4673}    \\
    \hline
  \end{tabular}
    \vspace{-0.3cm}
\end{table}

\begin{table}[htbp]
  \vspace{-0.3cm}
\footnotesize
  \caption{Results on the VoxCeleb1 dataset. LDE denotes learnable dictionary encoding \cite{cai2018exploring}, and SPE denotes spatial pyramid encoding \cite{jung2019spatial}}
  \label{tab:vox}
  \centering
  \begin{tabular}{p{3.8cm}<{\centering}p{0.98cm}<{\centering}p{0.8cm}<{\centering}p{0.8cm}<{\centering}}
    \hline
    System              &   Train set   &   EER(\%)     &   minDCF  \\
    \hline
    ResNet34-LDE \cite{cai2018exploring}   &   Vox1            &   4.56  &   0.441   \\
    ResNenXt41-MFA \cite{gao2019improving} &   Vox1            &   4.09  &   0.4578  \\
    ResNet34-SPE  \cite{jung2019spatial}  &   Vox1            &   4.03  &	0.4020  \\
    ResNet34-FPM \cite{jung2020improving}  &   Vox1            &   3.22  &	0.3500  \\
    ResNet34-BMFA-AFM(ours)&Vox1            &   \textbf{2.98}  &   \textbf{0.3008}  \\
    \hline
    ResNet50 \cite{chung2018voxceleb2}      &   Vox2            &   3.95  &	0.429   \\
    ThinResNet34-GhostVLAD \cite{xie2019utterance} &    Vox2   &	3.22  &   N/A \\
    ResNet34-SPE \cite{jung2019spatial}   &    Vox2           &	2.61  &	0.245   \\
    ResNet34-FPM \cite{jung2020improving}   &   Vox2            &	1.98	&   0.2050  \\
    ResNet34-BMFA-AFM (ours) &   Vox2        &	\textbf{1.73}  &  \textbf{0.1873}  \\
    \hline
  \end{tabular}
  \vspace{-0.5cm}
\end{table}
\section{Conclusions}
In this paper, we proposed a novel MFA method with bidirectional aggregation strategy and attentional fusion module for speaker verification. The bidirectional  aggregation  strategy combines and refines the feature maps from different stages iteratively in both a bottom-up and top-down manner for learning better speaker embedding. Applying the attentional fusion module can significantly improve the performance of BMFA by computing the optimal fusion weights for input feature maps, which is also effective for other MFA strategies. Experiments conducted on the SRE16 and VoxCeleb1 datasets demonstrate that our proposed method outperforms existing MFA methods for the SV task.
\section{Acknowledgements}
This work was partially funded by the National Natural Science Foundation of China (Grant No. U1836219).

\bibliographystyle{IEEEtran}

\bibliography{mybib}


\end{document}